\documentclass[twocolumn,aps,prl,showpacs,preprintnumbers,amsmath,amssymb]{revtex4}
\usepackage{graphicx}
\usepackage{dcolumn}
\usepackage{bm}
\begin{document}
\def\ibd {{\it ibid. }}
\def\bfq{{\bf q}}
\def\bfr{{\bf r}}
\def\bfk{{\bf k}}
\def\bfp{{\bf p}}
\def\eps{\epsilon}
\def\vep{\varepsilon}
\def\pl{\partial}
\def\tl{\tilde}
\def\ep{\epsilon}
\def\zbar{{\bar z}}
\def\al{\alpha}
\def\be{\beta}
\def\gam{\gamma}
\def\lam{\lambda}
\def\bfrho{\bf \rho}
\def\nab{\bf \nabla}
\def\Delmu{\Delta \mu}
\def\beq{\begin{equation}}
\def\eeq{\end{equation}}
\def\bea{\begin{eqnarray}}
\def\eea{\end{eqnarray}}
\title{Discontinuity in the specific heat  of a
weakly interacting Bose gas}
\author{Sang-Hoon \surname{Kim}\footnote{shkim@mmu.ac.kr}}
\affiliation{Division of Liberal Arts, Mokpo National Maritime University,
 Mokpo 530-729, Korea}
\date{\today}
\begin{abstract}
We produce the discontinuity in the specific heat of a homogeneous,
dilute, and weakly interacting Bose gas in a short-wavelength range
with a simple statistical method. 
The magnitude of the discontinuity
at the phase transition temperature
 is obtained as a function of the density and scattering length of the Bose particles.
\end{abstract}
\pacs{67.40.-w, 21.60.Fw, 03.75.Hh, 05.70.Fh}
\keywords{Bose system, Bose-Einstein condensation}
\maketitle


$^4$He becomes superfluid below 2.18K, so called $\lambda$ point, at
low atmospheric pressure. The specific heat of liquid $^4$He is
infinitesimal around 0$K$, and rises with $T^{2/3}$ until it reaches
a the $\lambda$ point that shows a dramatic divergence, and then
decreases tending asymptotically to a constant classical value.
 The transition is known as the strongest
degeneracy effect of a boson system, but it is still not fully
understood.
Many
physicists have tried to reproduce the discontinuity of the specific
since the discovery of the transition, but even until now very
little researches has done it successfully.

Kikuchi {\it et al.} applied  a partition function proposed by
Feynman to an Ising model-like 2D cubic lattice and considering
nearest-neighbors only, they calculated the specific heat at $T_c$
numerically \cite{kikuchi}. Although they produced the discontinuity
of the specific heat, their method strongly depends on various
atomic bond models and experimental variables. Therefore, this
semi-empirical and numerical method could not give any formula
between the $\lambda$ transition and the interaction strength.

The  objective of this paper is to reproduce the discontinuity of
the specific heat of a homogeneous, dilute, and weakly interacting
(HDWI) Bose system analytically. Through the application of a simple
statistical method. the relation between the $\lambda$ transition
and the interacting strength will be obtained. The weakly
interacting Bose model may prove useful to understand the strongly
interacting liquid $^4$He. We will write this argument in
D-dimensions ($ 2< D \le 3$) for better applicability.


The dilute is expressed by a dimensionless gas parameter $\gam =
n(1)a $, where $n(1)$ is the one-dimensional number density and $a$
is the $s$-wave scattering length. In a homogeneous system, the
density in D-dimensions is expressed as $n(D)=n(1)^D$.
 In a repulsive dilute gas, $\gam$ is positive and much less than 1.
The $\gam$ of liquid $^4$He is known as on the order of one-half,
and therefore the HDWI model is different from the real Bose system.

For a momentum-independent interaction $g(D)$, the mean field
contribution to the self-energy is $n(D) g(D)$. Then, the dispersion
relation of the weakly interacting Bose gas can be written as
\beq
\vep =  \vep_0 + n(D) g(D),
 \label{60}
\eeq
where $\vep_0=p^2/2m$.
The $g(D)$ is the positive coupling
constant in D-dimensions, where $2 < D \le 3$.
With the hard-sphere interaction,
it is well-known in 3D as
$ g(3) = 4 \pi \hbar^2 a/m.$
The dispersion relation corresponds to a short-wavelength range
in the Bogoliubov energy spectrum \cite{bogol}.
 The exact form of the
interaction is $2n(D)g(D)$ instead of $n(D)g(D)$ because the two
mean field contributions to the self-energy from Hartree and Fock
are equal. However, we will use Eq. (\ref{60}) because the factor of
2 is not a key concept here.

The grand partition function of the momentum-independent potential
is given from Eq. (\ref{60}) as
\bea
{\cal Q}(z,n,T)&=& \prod_\bfp\frac{1}{1-z e^{-\be (\vep_0(p)+n g)}}
\nonumber \\
&=& \prod_\bfp\frac{1}{1-z_e e^{-\be \vep_0(p)}},
\label{70}
\eea
where $\be=1/k_B T$. The $z$ is the  fugacity given by $z=e^{\be \mu}$
and 1 below $T_c$. The $\mu$ is the chemical potential, which is $0$
below the $T_c$ and negative above the $T_c$. The $z_e$ is the
effective fugacity given by
\beq
z_e = z e^{-\be n g } =  e^{-\be(|\mu|+n g)}.
\label{80}
\eeq
Note that  $0 \leq z_e < 1$.
As $T \rightarrow T_c$,
 then $z_e \rightarrow e^{ -\be_c n g} \equiv \eta_c $.
Note that $0< \eta_c<1$, and 1 for the non-interacting Bose system.
The effective chemical potential is written in the same way : $\mu_e
= \mu - ng$. It is $-ng $ instead of zero below $T_c$. Even for a
momentum-independent potential, it shifts the transition
temperature.

The equation of the state of the interacting system
in D-dimensions is written from
Eq. (\ref{70}) as \cite{huang}
 \beq n(D) \lam^D = g_{\frac{D}{2}}(z_e),
\label{90} \eeq where
$\lam=\sqrt{2\pi\hbar^2/mk_BT}$ is the
thermal wavelength, and $g_s(z)=\sum_{l=1}^{\infty} z^l /l^s$ is the
Bose function.
The only difference of the formula from the non-interacting model
 is the effective fugacity $z_e$,
and the information of the coupling constant $g(D)$ is in it.

Let $T_o \equiv 2\pi \hbar^2 n(2)/m k_B =1$
since $n(D)^{\frac{2}{D}}=n(2)$.
 It is about 5.92K for liquid $^4$He.
  In this way  from Eq. (\ref{90})
the relation of the transition temperature between
ideal and interacting system is written as
\beq
 T_c^0(D)^\frac{D}{2} g_{\frac{D}{2}}(1)
= T_c(D)^{\frac{D}{2}}
 g_{\frac{D}{2}}(\eta_c)=1.
\label{95} \eeq The right-hand side is actually $T_o$. In this way
the $T_c(3)$ of the interacting systems is obtained
self-consistently by numerical method  as \beq T_c(3) =
g_{\frac{3}{2}}(\eta_c(T_c))^{-\frac{2}{3}}. \label{100} \eeq The
relation between $T_c(3)$ and $T_c^0(3)$ is also written as the
power series of the $\gam$.
 The leading order in 3D is
\beq T_c(3)  \simeq  T_c^0(3)(1+ c \gam). \label{6} \eeq
We may use
the value obtained from the numerical simulation as
$c \simeq 1.3$ \cite{arnold,kashunikov}.


We obtain the internal energy in D-dimensions from
$U(T)=-(\partial/\partial\beta)_v  \ln {\cal Q}(z,n,T)$.
\\
At $T < T_c$ :
\\
 \beq
\frac{U^{-}(T)}{N k_B} = \frac{D}{2} T^{ \frac{D}{2} +1}
g_{\frac{D}{2} +1}(\eta),
\label{149}
\eeq
\\
and at $T > T_c$ :
\\
\beq
 \frac{U^{+}(T)}{N k_B} = \frac{D}{2} T^{ \frac{D}{2} +1} g_{
\frac{D}{2} +1}(z_e). \label{150} \eeq  The internal energy is
continous  at $T_c$ as $ \lim_{T \rightarrow T_c} \Delta U(T)
 =0$

\begin{figure}
\includegraphics{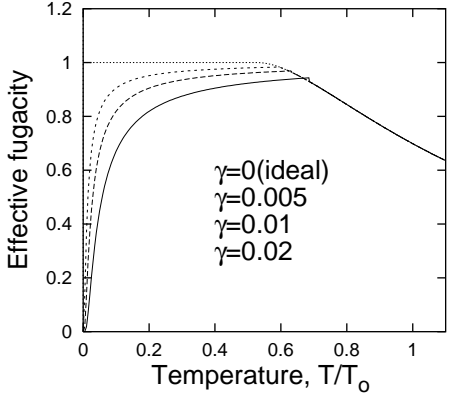}
\caption{ Effective fugacity of the weakly interacting Bose gas in
three-dimensions.}
\end{figure}
\begin{figure}
\includegraphics{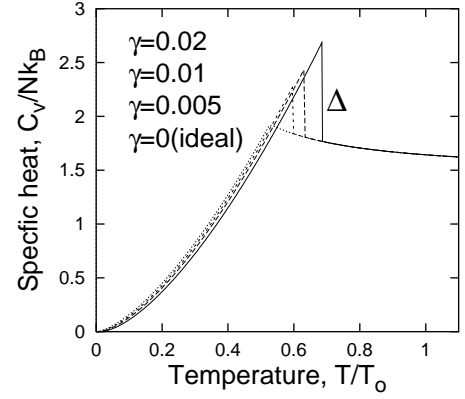}
\caption{Specific heat of the weakly
interacting Bose gas in three-dimensions.}
\end{figure}

The specific heat at constant volume is obtained as
$C_v = (\partial/\partial T)_v U(T)$.
 Therefore, we have $C_v$ below
and above the transition temperature in 3D.
\\
At $T < T_c$ :
\\
 \beq
\frac{C_v^{-}}{N k_B}
= \frac{15}{4} T^{ \frac{3}{2} } g_{ \frac{5}{2}}(\eta) +
3\gam  T^{\frac{1}{2} } g_{ \frac{3}{2}}(\eta),
\label{152}
\eeq
\\
and at $T > T_c$ :
\\
\beq
\frac{C_v^{+}}{N k_B} = \frac{15}{4}
  \frac{g_{ \frac{5}{2}}(z_e)}{g_{ \frac{3}{2} }(z_e)} -
\frac{9}{4}  \frac{g_{ \frac{3}{2} }(z_e)}
 {g_{ \frac{1}{2}}(z_e)}.
\label{153}
 \eeq
We used Eq. (\ref{95}) and the relation
 $ \be n(3) g(3) = 2 \gam/T$
and
$\partial \eta/\partial T = 2\gam\eta/T^2$.
The discontinuity of the $C_v$ at $T_c$ is
\beq \lim_{T
\rightarrow T_c}\frac{C_v^- - C_v^+}{N k_B} = 3\gam
T_c^{\frac{1}{2} }g_{ \frac{3}{2} }(\eta_c)
 +  \frac{9}{4}  \frac{g_{ \frac{3}{2} }(\eta_c)}
{g_{\frac{1}{2}}(\eta_c)}.
\label{200}
\eeq
 The non-interacting limit
goes to the well-known results as $ \lim_{\gam \rightarrow 0} \Delta
C_v =  0$, since $g_{\frac{1}{2}}(1)$ goes to infinity.

The effective fugacity $z_e(T)=z
e^{-\eta}$ is shown in Fig. 1.
It downshifts from 1 as the interaction is increasing.
The $C_v$ at $D=3$ is shown in Fig. 2, too.


We considered the short-wavelength range only in the dispersion
relation, and applied a simple mean field based statistical
mechanics to the homogeneous, dilute, and weakly interacting Bose
system. In spite of the simplicity of the system, it successfully
produced the relation between the discontinuity of the specific heat
and the interaction strength of the model clearly.


{}
\end{document}